# An extended low-frequency noise compact model for single-layer graphene FETs including correlated mobility fluctuations effect

Nikolaos Mavredakis, Anibal Pacheco-Sanchez, and David Jiménez

*Abstract*— Correlated mobility fluctuations are considered in the physics-based carrier number fluctuation (*ΔN*) low-frequency noise (*LFN*) compact model of single-layer graphene field effect transistors (GFET) in the present study. Trapped charge density and Coulomb scattering coefficient *ΔN LFN* parameters are obtained after applying a parameter extraction methodology, adapted from conventional silicon technologies, to the linear ambipolar regions of GFETs. Appropriate adjustments are considered in the method according to GFETs' physical characteristics. Afterwards, Hooge mobility as well as series resistance fluctuations *LFN* parameters can be extracted. The updated *LFN* model is validated with experimental data from various long and short-channel GFETs at an extended range of gate and drain bias conditions.

## I. Introduction

After two decades of extensive research since graphene's discovery, graphene field-effect transistor (GFET)-based applications are approaching rapidly the commercialization stage. Experimental 2D-technology pilot lines have been contributing to the latter through targeting large scale on wafer integration [1]-[3]. Due to graphene's exceptional low-field mobilities *μ*, saturation velocities *u*$_{sat}$, which enable impressive *RF* performance (cut-off frequencies approaching *500 GHz* and maximum oscillation frequencies around *200 GHz*), as well as its unique ambipolar transport characteristic, GFETs are recognized as strong contenders for multifunctional analog/*RF* circuit-design, approaching the performance of established CMOS and III–V technologies despite the early stage of GFET development [4]–[6]. Such circuits are prone, up to some extent, to low-frequency noise (*LFN*) due to *LFN* up-conversion to phase noise in frequency multipliers [7] and oscillators [5], [8], for example. Besides, industry-based GFETs have already been employed in biosensors targeting COVID [9], [10] and cancer [11] detection with excellent response while great advances have been recorded lately in GFET neuroelectronics such as in brain neural sensing platforms mainly via solution-gated (SG) GFETs [12], [13]. *LFN* defines the detection limit as well as the long-term functionality of such sensors [13], as it degrades the signal quality [12]. Additionally, GFET-based *THz* detection applications have been proposed with promising results due to unique opto-electronic properties of graphene, offering competitive performance in broadband, room-temperature operation and providing an alternative to silicon and III–V technologies for imaging spectroscopy and medical applications [14]. Thermal noise is only considered in such detectors usually [15], but *LFN* can also be critical as it sets the lower limit for the frequency modulation required in these applications [15].

*LFN* has been thoroughly investigated in GFETs [16]-[24], mainly recorded to be inversely proportional to frequency (~*1/f*). Carrier number due to trapping/detrapping (*ΔN*), empirical Hooge mobility (*Δμ*) and series resistance (*ΔR*) fluctuations have been found as the main GFET *LFN* generators, similar to other FET types. A *ΔN*-induced M-shape gate voltage ($V_{GS}$) dependence of output noise divided by squared drain current ($S_{ID}/I_D^2$) with a minimum at charge neutrality point (*CNP*), has been experimentally recorded and analyzed [16]-[21], [23] while *Δμ* and *ΔR* mechanisms have been shown to contribute primarily at *CNP* (Λ-shape *Δμ* with a maximum at *CNP*) and at high $I_D$ (*ΔR*), respectively [16], [19]-[21]. Besides, high-field effects at increased drain voltage ($V_{DS}$) result in a decrease of *LFN* [19]-[22]. For fast simulations of *LFN*-sensitive GFET integrated circuits, compact models are essential to accurately predict the behavior of devices and circuits. A physics-based single-layer (SL) GFET *LFN* compact model has been proposed by our group [19]-[21, §5] and has been experimentally validated for both lab- and industry-fabricated GFETs; extending such model for multilayer devices entails addressing interlayer coupling and screening effects, which

This work has received funding from the European Union's Horizon 2020 research and innovation programme under grant agreements No GrapheneCore3 881603, from Ministerio de Ciencia, Innovación y Universidades under grant agreements RTI2018-097876-B-C21(MCIU/AEI/FEDER, UE), PID2021-127840NB-I00 (MCIN/AEI/FEDER, UE) and CNS2023-143727 RECAMBIO (MCIN/AEI/10.13039/501100011033). This work is also supported by the European Union NextGenerationEU/PRTR research project.
N. Mavredakis, and D. Jiménez are with the Departament d'Enginyeria Electrònica, Escola d'Enginyeria, Universitat Autònoma de Barcelona, Bellaterra 08193, Spain. (e-mail: Nikolaos.mavredakis@uab.cat).
A. Pacheco is with the Departamento de Electrónica y Tecnología de Computadores, Universidad de Granada, 18011 Granada, Spain.

is a future work task. Recently, *LFN* modeling updates have been proposed [24], where a more accurate *CNP* square root approximation of quantum capacitance $C_q(=k\sqrt{V_c^2+C_1^2})$ is adapted; $k$ is a well-defined constant of the chemical potential ($V_c$)-based *IV* model, $C_1=U_T.ln(4)$ with $U_T(=25.6\ mV)$ the thermal voltage at *300 K* [21, §2.1.2]. Besides, mobility reduction due to vertical field through $\theta_{int}$ parameter [25], [26], ambipolarity [27] and a more efficient constant velocity saturation (*VS*) model, have been considered in [24]. Importantly, the complete GFET model [21] is modular and has already been extensively validated experimentally from *DC* up to the *RF* regime, including non-quasi-static effects, small-signal analysis, and high-frequency noise among others [21], ensuring that the *LFN* module is consistently integrated within the complete modeling framework.

No other such complete GFET *LFN* model is available in literature than the one proposed by our group [21, §5], and what is still missing, is the well-known correlated mobility fluctuation effect on *ΔN LFN* (*ΔN+α_C*) with $α_C$ the Coulomb scattering coefficient in *V.s/C* [28], [29]. *ΔN+α_C* has been referred to affect *LFN* in GFETs [16], [18] but has not been included in a compact model so far. *LFN* mechanisms have been thoroughly investigated through experimental characterization and modelling in incumbent CMOS technologies [28]-[34] where complete models have been proposed in [28] (*ΔN+α_C*) and [30, §6] (*ΔN+α_C, Δμ, ΔR*). A simple $S_{ID}/I_D^2 \sim g_m/I_D$ (*ΔN+α_C*) model has been employed in [29], [31], valid in linear region; $g_m$ is the measured transconductance [31, (1)-(7)]. Parameter extraction methods regarding this $\sim g_m/I_D$ model have been proposed in [32] and rectified in [33], [34] to account for series resistance $R_c$ effects on $g_m$. Until now *ΔN+α_C* has not been included in our model, and its impact has been captured empirically through increased $N_t$ and Hooge $α_H$ parameters [19]-[21], [24]. Hence, there is a necessity to distinguish such contributions for better physical representation of the model. Here, we propose for the first time an extended *LFN* model that explicitly incorporates correlated mobility fluctuations, yielding more physically correct parameter values (avoiding the previous overestimation of $α_H$ and $N_t$ due to neglecting $α_C$ [19]-[21], [24]) and ensuring the physical validity of the model, while also providing improved scalability for testing multiple devices in industrial processes. The proposed model assumes a uniform spatial distribution of traps near the dielectric interface, which is valid for the relatively large channel dimensions of the GFETs under study [28]. This uniformity leads to the typical *1/f* trend, as no Lorentzian-like *LFN* spectra, characteristic of few-trap systems, have been experimentally observed so far. Note that $I_D$ variability and *1/f* noise have been proved to be identical mechanisms for CMOS [35], [36], organic FETs [37], [38] and GFETs [39]. Recently, we have proposed a (spatial) $I_D$ variance model for GFETs [39] with excellent model-measurements agreements where *ΔN+α_C* is the dominant physical mechanism. $I_D$ variance is generated by deviations of both graphene charge $δQ_{gr}$ and effective mobility $δμ_{ueff}$ induced by charged impurities fluctuations $δQ_{imp}$ locally in the channel [39]. Here, we follow the $I_D$ variance model formulations [39] to describe (temporal) *ΔN+α_C LFN*, where $δQ_{gr}, δμ_{ueff}$ are now generated by charged traps fluctuations $δQ_t$, and not by $δQ_{imp}$ as in $I_D$ variance case ($Q_t$ is included in $Q_{imp}$). A $g_m/I_D$-based *ΔN+α_C LFN* parameter extraction procedure, adapted from [32], is adjusted appropriately to GFETs to avoid extrinsic $R_c$ effects, while a path is proposed to accurately account for potential *Δμ* effects near *CNP*. We also extract *ΔR*-related parameters accordingly.

This work introduces the first compact GFET *LFN* model explicitly including correlated mobility fluctuations. Its formulation follows the established $I_D$ variance model, reflecting the equivalence of the underlying mechanisms, while accounting for the different nature of the parameters [39]. A structured sequential parameter extraction procedure, applied separately to p(n)-type operation, ensures convergence. The model is validated on diverse liquid-gated and *RF* short-channel GFETs, across extended $V_{GS}$ and $V_{DS}$ ranges, and for both input- and output-referred *LFN*.

## II. DEVICES AND MODEL DERIVATIONS

*IV* and *LFN* on-wafer experimental data from different types of SL GFETs are employed to verify the proposed parameter extraction methodology and eventually to validate the updated *LFN* module. Using devices from different technologies and spanning a wide range of geometries enables assessing the robustness and scalability of the proposed *LFN* model. In more detail, i) long-channel top-gated SG SL GFETs with *W/L=100/100, 50/50* and *20/20 μm/μm*, where *W, L* are the channel width and length, respectively [12], [39], as well as ii) type A, B, C, D CVD grown short-channel back-gated $Al_2O_3$ (~*4 nm*) SL *RF* GFETs [19], [20] have been measured for an extended range of $V_{GS}$ sweeps, including both strong p(n)-type regions and *CNP*, for $V_{DS}$=*50 mV* in (i) and $V_{DS}$=*30m, 60m, 0.1, 0.2, 0.3 V* in (ii) (to consider high-field *VS* effect on *LFN*). Measured *LFN* spectra from *1 Hz* to *1 kHz* are examined. Although SG long-channel GFETs lack a physical top-gate dielectric, they exhibit a measured top-gate capacitance of ~*2 μF/cm²*, equivalent to an oxide thickness of *1.7 nm*, which is used for the modeling purposes in this work. Short A–D GFETs are located at different dies of the same wafer, and the prototype-level maturity of the technology leads to die-to-die variability, mainly in $μ$ and residual charge, as reflected in their *IV* characteristics in [19], [20]. Further details on the fabrication and measurements setups of the devices under test (DUT) can be found elsewhere [12], [19], [20], [39].

The *LFN* model derivations are formulated by considering uncorrelated local noise fluctuations at microscopic $Δx$ slices of the channel which induce voltage or current variations. By integrating such deviations from Source (S) to Drain (D), total noise Power Spectral Density (PSD) is estimated [19]-[21, §5], [30, §6]:

$$S_{ID} = \int_0^L G_{CH}^2 \delta R^2 \frac{S_{\delta I_n^2}}{\Delta x} dx \qquad (1)$$

with $S_{\delta I_n^2}$ the local noise PSD and $G_{CH}, \delta R$ the local channel conductance and resistance, respectively, where $G_{CH}\delta R=\Delta x/L$ [39, (1)]. After considering both charge conservation law and fundamental *IV* GFET model principles [39, (2)-(8)], the relative current fluctuation due to $\Delta N+\alpha_C$ ($\Delta N$ from now on) effect equals to:

$$\frac{\delta I_x}{I_D} = \frac{\partial Q_{gr(x)}}{Q_{gr(x)}}\Big|\delta Q_t \pm \frac{\partial\mu_{ueff}}{\mu_{ueff}}\Big|\delta Q_t = \delta Q_t\left(\frac{k|V_c|}{(k|V_c|+C)Q_{gr(x)}} \pm \alpha_c\mu_{ueff}\right) \quad (2)$$

TABLE I- *LFN* M<small>ODEL</small> E<small>QUATIONS</small>

| |
|---|
| $\Delta NA_a = \frac{4L_{eff}}{Cg_{vc}} \left\{ \frac{1}{\varphi_1}\left[\frac{\mp C^3}{k} \mp C^2\gamma_1\right] + \frac{1}{\varphi_2}\left[\frac{C^2}{k}\tanh^{-1}\left(\frac{V_c}{\gamma_1}\right) \pm \frac{C^3}{k\gamma_2}\tanh^{-1}\left(\frac{\gamma_2}{\gamma_1\gamma_3}\right)\right] + \frac{1}{\varphi_3}\left[-\frac{\sqrt{\alpha}C\gamma_4}{\sqrt{k}}\tan^{-1}\left(\sqrt{\frac{k}{\alpha}}V_c\right) \mp 2\alpha C\sqrt{k}\gamma_5\tan^{-1}\left(\frac{\sqrt{k}\gamma_1}{\gamma_5}\right) + 2\alpha C^2\tanh^{-1}\left(\frac{V_c}{\gamma_1}\right) - \alpha\gamma_3\tanh^{-1}\left(\frac{V_c}{\gamma_1}\right) + \sqrt{\alpha}\gamma_4\gamma_5\tanh^{-1}\left(\frac{V_c\gamma_5}{\sqrt{\alpha}\gamma_1}\right) \pm 2\alpha C\gamma_2\tanh^{-1}\left(\frac{\gamma_2}{\gamma_1\gamma_3}\right) \mp \alpha C^2\ln\left(\frac{(C\pm kV_c)^2}{\alpha+kV_c^2}\right)\right]\right\}\Big|_{V_{cd}}^{V_{cs}}$ (6a) |
| $\Delta NA_b = \frac{4\mu_u}{Cu_{sat}}\left\{\frac{1}{\varphi_3}\left[\frac{\mp C^2k^2\gamma_1}{C\pm kV_c} \mp \alpha Ck^2\frac{\gamma_1}{\gamma_6} - \frac{k^2V_c\gamma_1\gamma_3}{2\gamma_6} \pm \frac{\alpha Ck^{\frac{3}{2}}}{\gamma_5}\tan^{-1}\left(\frac{\sqrt{k}\gamma_1}{\gamma_5}\right) \pm C^2k\tanh^{-1}\left(\frac{V_c}{\gamma_1}\right) - \frac{b^2k^2\gamma_4}{2\sqrt{\alpha}\gamma_5}\tanh^{-1}\left(\frac{V_c\gamma_5}{\sqrt{\alpha}\gamma_1}\right) \pm \frac{C^3k}{\gamma_3}\tanh^{-1}\left(\frac{\gamma_2}{\gamma_3\gamma_1}\right)\right] + \frac{1}{\varphi_4}\left[\pm 2Ck^{\frac{3}{2}}\gamma_4\gamma_5\tan^{-1}\left(\frac{\sqrt{k}\gamma_1}{\gamma_5}\right) + C^2k\gamma_7\tanh^{-1}\left(\frac{V_c}{\gamma_1}\right) - 2C^2k\gamma_4\tanh^{-1}\left(\frac{V_c}{\gamma_1}\right) - \frac{C^2k\gamma_5\gamma_7}{\sqrt{\alpha}}\tanh^{-1}\left(\frac{V_c\gamma_5}{\sqrt{\alpha}\gamma_1}\right) + 2Ck\gamma_3\tanh^{-1}\left(\frac{\gamma_2}{\gamma_3\gamma_1}\right)\right]\right\}\Big|_{V_{cs}}^{V_{cd}}$ (6b) |
| $\Delta NB_a = \frac{L_{eff}(\alpha_c\mu_{ueff})^2}{Cg_{vc}}\left\{\frac{V_c\gamma_1(4\alpha+k\gamma_8)}{8} - \frac{C_1^2(C_1^2k-4\alpha)}{8}\ln(\gamma_1+V_c) + \frac{\alpha CV_c}{k} + \frac{CV_c^3}{3}\right\}_{V_{cd}}^{V_{cs}}$ (7a)   $\Delta NB_b = \frac{(\alpha_c\mu_{ueff})^2\mu_uk}{Cu_{sat}}\left\{\frac{V_c\gamma_1+C_1^2\tanh^{-1}\left(\frac{V_c}{\gamma_1}\right)}{2}\right\}_{V_{cs}}^{V_{cd}}$ (7b) |
| $\Delta NC_a = \frac{2\alpha_c\mu_{ueff}L_{eff}}{k^2Cg_{vc}}\{2Ck|V_c| \mp 2Ck\gamma_1 + k^2V_c\gamma_1 \mp 2C(C+\gamma_3)\ln(C\pm kV_c) + (2C^2+C_1^2k^2)\ln(V_c+\gamma_1) \pm 2C\gamma_3\ln(\gamma_2+\gamma_3\gamma_1)\}_{V_{cd}}^{V_{cs}}$ (8a) |
| $\Delta NC_b = \frac{4k\alpha_c\mu_{ueff}\mu_u}{Cu_{sat}\varphi_2}\left\{\mp\frac{C\gamma_5\sqrt{\varphi_2}}{\sqrt{k}}\tan^{-1}\left(\frac{\sqrt{k}\gamma_1}{\gamma_5}\right) + \alpha\tanh^{-1}\left(\frac{V_c}{\gamma_1}\right) + \frac{C^2}{k}\tanh^{-1}\left(\frac{V_c}{\gamma_1}\right) - \sqrt{\alpha}\gamma_5\tanh^{-1}\left(\frac{V_c\gamma_5}{\sqrt{\alpha}\gamma_1}\right) \pm C\gamma_3\tanh^{-1}\left(\frac{\gamma_3}{\gamma_3\gamma_1}\right)\right\}_{V_{cs}}^{V_{cd}}$ (8b) |
| $\Delta\mu_a = \frac{L_{eff}}{kg_{vc}}\left\{CV_c + \frac{k}{2}V_c\gamma_1 + k\frac{C_1^2}{2}\ln[V_c+\gamma_1]\right\}_{V_{cd}}^{V_{cs}}$ (9a)   $\Delta\mu_b = \frac{\mu_u}{u_{sat}}\left\{\tanh^{-1}\left(\frac{V_c}{\gamma_1}\right) - \frac{\gamma_5\tanh^{-1}\left(\frac{\gamma_5V_c}{\sqrt{\alpha}\gamma_1}\right)}{\sqrt{\alpha}}\right\}_{V_{cs}}^{V_{cd}}$ (9b) |
| with $\varphi_1=\gamma_2(C\pm kV_c)$, $\varphi_2=\alpha k+C^2$, $\varphi_3=\varphi_2^2$, $\varphi_4=\varphi_3^2$, and $\gamma_1=\sqrt{(V_c^2+C_1^2)}$, $\gamma_2=C_1^2k\mp CV_c$, $\gamma_3=\sqrt{(C^2+C_1^2k^2)}$, $\gamma_4=\sqrt{(C^2-\alpha k)}$, $\gamma_5=\sqrt{(\alpha-C_1^2k)}$, $\gamma_6=\alpha+kV_c^2$, $\gamma_7=C^2-3\alpha k$, $\gamma_8=2V_c^2+C_1^2$ and $V_{cs(d)}$ is $V_c$ at S(D). $\pm, \mp$: top sign refers to $V_c>0$ and bottom to $V_c<0$. |

$C=C_t+C_b$ where $C_{t(b)}$ is the top (back) gate oxide capacitance. By considering (1), (2), $\Delta N\ S_{ID}/I_D^2$ yields [39, (9)]:

$$\frac{S_{ID}}{I_D^2}\Big|\Delta N = \frac{e^2N_t}{L^2W}\int_0^L\left(\frac{k|V_c|}{(k|V_c|+C)Q_{gr(x)}} \pm \alpha_c\mu_{ueff}\right)^2 dx \quad (3)$$

with $e$ the electron charge. Trapped charge density $N_t$ (in $cm^{-2}$) [21, §5], [24, 30, §6] and $\alpha_C$ are used as $\Delta N$ *LFN* model parameters. Notice that $\alpha_C$ has been recorded $V_{GS}$-independent in SL GFETs [18], [39], in contrast to CMOS where a weak dependency has been reported [28, Fig. 3]. Following an identical methodology (as in $\Delta N$ case), $\Delta\mu\ S_{ID}/I_D^2$ is derived as [19], [20], [24], [30, §6]:

$$\frac{S_{ID}}{I_D^2}\Big|\Delta\mu = \frac{\alpha_H e}{WL^2}\int_0^L\frac{1}{Q_{gr(x)}}dx \quad (4)$$

where $\alpha_H$ is the unitless Hooge model parameter. $\Delta\mu$ stems from phonon scattering [31] and $\alpha_H$ is considered a technology dependent parameter, which maximizes when phonon scattering dominates over Coulomb scattering ($\alpha_C$) [31, (10), Fig. 2b]. Finally, $\Delta R\ S_{ID}/I_D^2$ is provided by a well-established CMOS model [31, (12)]:

$$\frac{S_{ID}}{I_D^2}\Big|\Delta R = s_{\Delta R}\left(\frac{g_m}{2} + g_{ds}\right)^2 \quad (5)$$

with $g_{ds}$ the extrinsic transconductance and $S_{\Delta R}$ ($\Omega^2/Hz$) the $R_c$ fluctuation also used as a model parameter. The sum of the distinct effects in (3)-(5) results in the total $S_{ID}/I_D^2$. The well-established $\sim 1/(WL)\ S_{ID}/I_D^2$ *LFN* dependence in CMOS [28]-[32], is also preserved in GFETs for both $\Delta N$, $\Delta\mu$ modules ((3)-(4)) while, as shown in (5), $\Delta R$ presents a $\sim(W/L)^2$ trend, following $g_m^2$ and $g_{ds}^2$ [21, (25)].

To successfully integrate the *LFN* module into the $V_c$-based (Verilog-A) model, the final closed-form compact equations must be also $V_c$-dependent. Hence, the integral variable in (3), (4) is altered from $dx$ to $dV_c$ [39, (10)]. Besides, $Q_{gr}(=k/2(V_c^2+\alpha/k))$ is also expressed in terms of $V_c$ [39, (3)]; $\alpha$ is a residual charge-related variable [19]-[21]. Thus, (6a–b), (7a–b) and (8a–b) ($S_{ID}/I_D^2|_{\Delta N}=e^2N_t/(L^2W)(\Delta NA+\Delta NB+\Delta NC)=e^2N_t/(L^2W)(\Delta NA_a-\Delta NA_b+\Delta NB_a-\Delta NB_b+\Delta NC_a-\Delta NC_b))$ and (9a–b) ($S_{ID}/I_D^2|_{\Delta\mu}=2e\alpha_H/(CL^2W)(\Delta\mu_a-\Delta\mu_b))$ [24, (12)-(16)], [39, (11)-(17)] are derived (in Table I) where $\Delta N$ generated only by $\delta Q_t$ is denoted by $\Delta NA$ ($\alpha_C=0$) while $\Delta NB+\Delta NC$ quantify the additional $\delta\mu_{ueff}$ contribution. The *VS*-induced decrease of *LFN* [19]-[20] is clearly identified in (6b)-(9b) (multiplied by $1/u_{sat}$). $L_{eff}$ is the effective channel length due to *VS* and $g_{vc}$ a normalized $I_D$ coefficient [21, §2.1.2], [39] while $\mu_u=\mu/(1+\theta_{int}\sqrt{(V_o^2+(V_{GS}-V_{G0})^2)})$ is the intrinsic (no $R_c$ effect) $V_{GS}$-dependent mobility degraded due to vertical field with $V_{o(G0)}$ the residual charge-related (overdrive gate) voltage [25], [26]; $\mu_{ueff}(=\mu_u/(1+|E_x|/E_C))$ with $E_x$ and $E_C$ the horizontal and critical electric fields, respectively, includes *VS* effects. Ambipolarity feature has already been introduced to *IV* [27], $I_D$ variance [39] and prior *LFN* modules [24] optimizing the model performance. Here, the proposed parameter extraction methodology is applied separately in p(n)-type regions resulting in distinct $N_{tp(n)}$, $\alpha_{Cp(n)}$, $S_{\Delta Rp(n)}$ extracted parameters. Note that a unique $\alpha_H$ value is extracted as it can only be critical around *CNP*.

One of the key advantages of a physics-based compact modular model is that it can be improved by incorporating a more solid physical foundation, as demonstrated in this work, without significantly affecting its usability at the circuit level. In particular, a more accurate representation of physical mechanisms such as the inclusion of correlated mobility fluctuations in $\Delta N$, enables more reliable performance projections of these effects at both the device and circuit levels. At the circuit level, this refinement could enable a more accurate assessment of the impact of each fluctuation mechanism on upconverted phase noise in nonlinear applications, such as mixers and frequency multipliers. Circuit-level simulations to quantify the influence of LFN on phase noise in RF circuits and on biosensor sensitivity are currently under development and will be addressed in future research. Nevertheless, the other model modules have already been accurately benchmarked in various circuit topologies, including biosensors [9]–[11] and analog/RF circuits, such as key wireless communication components like subharmonic mixers, phase shifters, power amplifiers, and modulators [6].

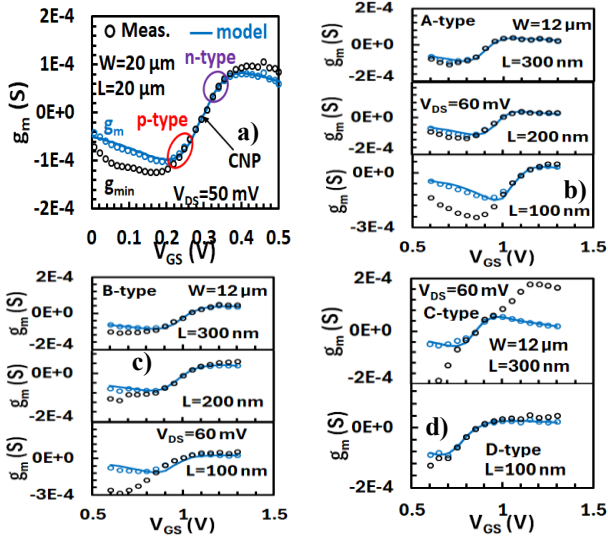

Fig.1. Transconductance $g_m$ vs. gate voltage $V_{GS}$ for a) solution-gated (SG) GFETs with width $W=20\,\mu m$ and length $L=20\,\mu m$, and for type A (b), B (c), C (upper panel of d) and D (bottom panel of d) RF GFETs with $W=12\,\mu m$ and $L=300$ (top panel), $200$ (middle panel) and $100$ (bottom panel) $nm$ at $V_{DS}=50$ (a) and $60$ (b-d) $mV$. Markers: measurements, solid lines: model. Large circles in (a) depict the regions where the method is applicable.

## III. Parameter Extraction Methodology

A $\sim g_m/I_D$ linear region ($\mu_{ueff}\approx\mu_u$) $\Delta N$ model is [29], [31]:

$$\frac{S_{I_D}}{I_D^2}|\Delta N = S_{VFB}\left(\frac{g_m}{I_D}\right)^2\left(1+\alpha_C\mu_u C \frac{I_D}{g_m}\right)^2 \quad (10)$$

where $S_{VFB}(=e^2 N_t/(WLC^2))$ is the flat-band voltage PSD. $S_{VG}(=S_{ID}/g_m^2)$ is the input PSD and $\sqrt{S_{VG}}$ is given by (4) of [32], which is adjusted accordingly for GFETs as:

$$\sqrt{S_{VG}}|\Delta N = \sqrt{S_{VFB}}\left(1+\alpha_C \frac{\mu}{1+\theta_{int}\sqrt{V_o^2+V_{Geff}^2}} C \frac{I_D}{g_m}\right) \quad (11)$$

where $V_{Geff}=V_{GS}-V_{G0}$ [26], [39]. (11) denotes a linear relation ($y=b_1 x+b_2$) between $y=\sqrt{S_{VG|\Delta N}}$ and $x=I_D/[g_m(1+\theta_{int}\sqrt{(V_o^2+V_{Geff}^2)})]=I_D/g_m\cdot\mu_u/\mu$ with $b_1=\sqrt{S_{VFB}}\alpha_C\mu C$ and $b_2=\sqrt{S_{VFB}}$. Hence, $\Delta N$ parameters can be straightforwardly extracted from $b_2$ from the intercept with y-axis (i.e., $N_t$) and $b_1$ from the slope (i.e., $\alpha_C$); $W$, $L$, $C$ are known while the extracted IV model parameters ($\mu$, $\theta_{int}$, $V_{G0}$, $V_o$) have been presented elsewhere [24], [26], [27], [39]. This methodology has been applied previously to MOSFET technologies with the corresponding underlying $\mu$ and $I_D$ models [32] (with $x=I_D/g_m$ in contrast to our case) where it is valid for long $L$ where $R_c$ effect on IV part is negligible. For short $L$ though, $R_c$ contribution (through extrinsic $\mu_u$ and $V_{DS}$ [34, (1)]), can decay the accuracy; ways to deal with this have been proposed in [33], [34]. In our GFET IV model, $R_c$ is not considered in $\mu_u$ as it is incorporated by connecting $R_c(=R_D=R_S)$ between intrinsic and extrinsic D, S, respectively, in the Verilog-A code [27], [39], thus $\theta_{int}$ is employed in (11) instead of $\theta_1$ in [32] ($\theta_1=\theta_{int}+2\mu CW/LR_C$ [26]). Hence, (11) accounts only for intrinsic channel-induced $\Delta N$ provided that it is applied in $g_m\approx g_{min}$ region (with $g_{min}$ the intrinsic transconductance calculated by the IV model [21]), where channel dominates over contact effects (cf. [34, (10)]). Regarding $\Delta\mu$, in contrast to CMOS processes where it emerges only at deep weak inversion [30, Fig. 6.12] and can be neglected in other regions due to exponential $I_D$ increase, ambipolar GFETs present a limited $I_D$ variation (<one order of magnitude) from CNP to strong p(n)-type regions [21], [26]-[27], [39], hence $\Delta\mu$ should be considered in our method, as it might contribute around CNP [19]-[21], [24]. Fig. 1 depicts remarkable measurements (markers)-model (lines) $g_m$ agreements for all DUTs. $g_{min}$ is also shown so as to identify $g_m\approx g_{min}$ regions where $\Delta N+\Delta\mu$ dominates and hence, the proposed method is valid (cf. large red and purple circles for both p- and n-type regions in Fig. 1a). Initially $N_t$, $\alpha_C$ are extracted from linear fits of (11), as described above, followed by the estimation of $\alpha_H$ parameter by fitting the complete channel LFN model [24] with measurements at CNP [20]. Since a potential strong $\Delta\mu$ effect can also affect the channel LFN not only at the CNP but also in its vicinity (within $g_m\approx g_{min}$), the simulated $S_{VG|\Delta\mu}$ (after $\alpha_H$ calculation) must be subtracted from the measured $S_{VG}$ as $S_{VG}=S_{VG|\Delta N}+S_{VG|\Delta\mu}+S_{VG|\Delta R}\leftrightarrow S_{VG|\Delta N}\approx S_{VG}-S_{VG|\Delta\mu}$ ($S_{VG|\Delta R}$ remains negligible when $g_m\approx g_{min}$) and (11) accounts only for $\Delta N$ effect. The linear-fit approach in (11) is then re-applied, in case of a non-negligible $\alpha_H$, to tune the $\Delta N$ parameters ($N_t$, $\alpha_C$), after which the full channel LFN model ($\Delta N+\Delta\mu$) is again simulated to verify its validity from CNP to the limits of the $g_m\approx g_{min}$ region. This procedure is repeated until an optimum model-measurements agreement is achieved across the entire region of interest without further changes to the channel LFN parameters. The methodology has been applied at both p(n)-type regions for all the DUTs, and the resulting linear fits are depicted in Fig. 2. $S_{\Delta R}$ parameter is then derived by adjusting the LFN model to capture the measurements at high $I_D$, where $\Delta N$, $\Delta\mu$ are minimal [19. Fig. 5c].

## IV. Results-Discussion

The extracted parameters are shown in Table II where $\Delta\mu$ presents a more notable contribution (around CNP through increased $\alpha_H$) for DUTs where $\alpha_C$ is less significant implying comparable phonon (which generates $\Delta\mu$ [31]) to Coulomb scattering. For the short-channel RF DUTs, $\mu$, which is mainly defined by Coulomb scattering mobility ($\sim 1/\alpha_C$) at low fields in GFETs [39], is maximum at the shorter $L$ [24],

TABLE II - *LFN* PARAMETERS

| Parameter | Units | SG20x20 | SG50x50 | SG100x100 | A100 | A200 | A300 | B100 | B200 | B300 | D100 | C300 |
|---|---|---|---|---|---|---|---|---|---|---|---|---|
| $N_{tn}$ (x10$^8$) | $cm^{-2}$ | 8.5 | 3.48 | 36 | 5.5 | 12.4 | 36 | 10.6 | 7 | 65 | 25.6 | 0.0815 |
| $N_{tp}$ (x10$^8$) | $cm^{-2}$ | 8.8 | 2.6 | 6.25 | 1.06 | 0.9 | 25.3 | 1.28 | 0.395 | 3.71 | 6.34 | 1.6 |
| $\alpha_{Cn}$ | $kV\,s/C$ | 1.8 | 4.5 | 3 | 30 | 110 | 85 | 60 | 136 | 50 | 100 | 32.6 |
| $\alpha_{Cp}$ | $kV\,s/C$ | 1.25 | 4.5 | 5 | 25 | 200 | 50 | 60 | 400 | 110 | 74 | 7.8 |
| $\alpha_H$ (x10$^{-5}$) | - | 0.7 | 0.7 | 0.7 | 3.4 | 0.1 | 0.1 | 45 | 0.1 | 0.1 | 40 | 30 |
| $S_{\Delta Rn}$ (x10$^{-4}$) | $\Omega^2/Hz$ | 60 | 12.2 | 19.1 | 0.01 | 400 | 600 | 0.01 | 0.01 | 0.01 | 150 | 17.15 |
| $S_{\Delta Rp}$ (x10$^{-4}$) | $\Omega^2/Hz$ | 3.35 | 0.01 | 0.232 | 6.5 | 50 | 200 | 20 | 0.01 | 0.01 | 0.01 | 3.5 |

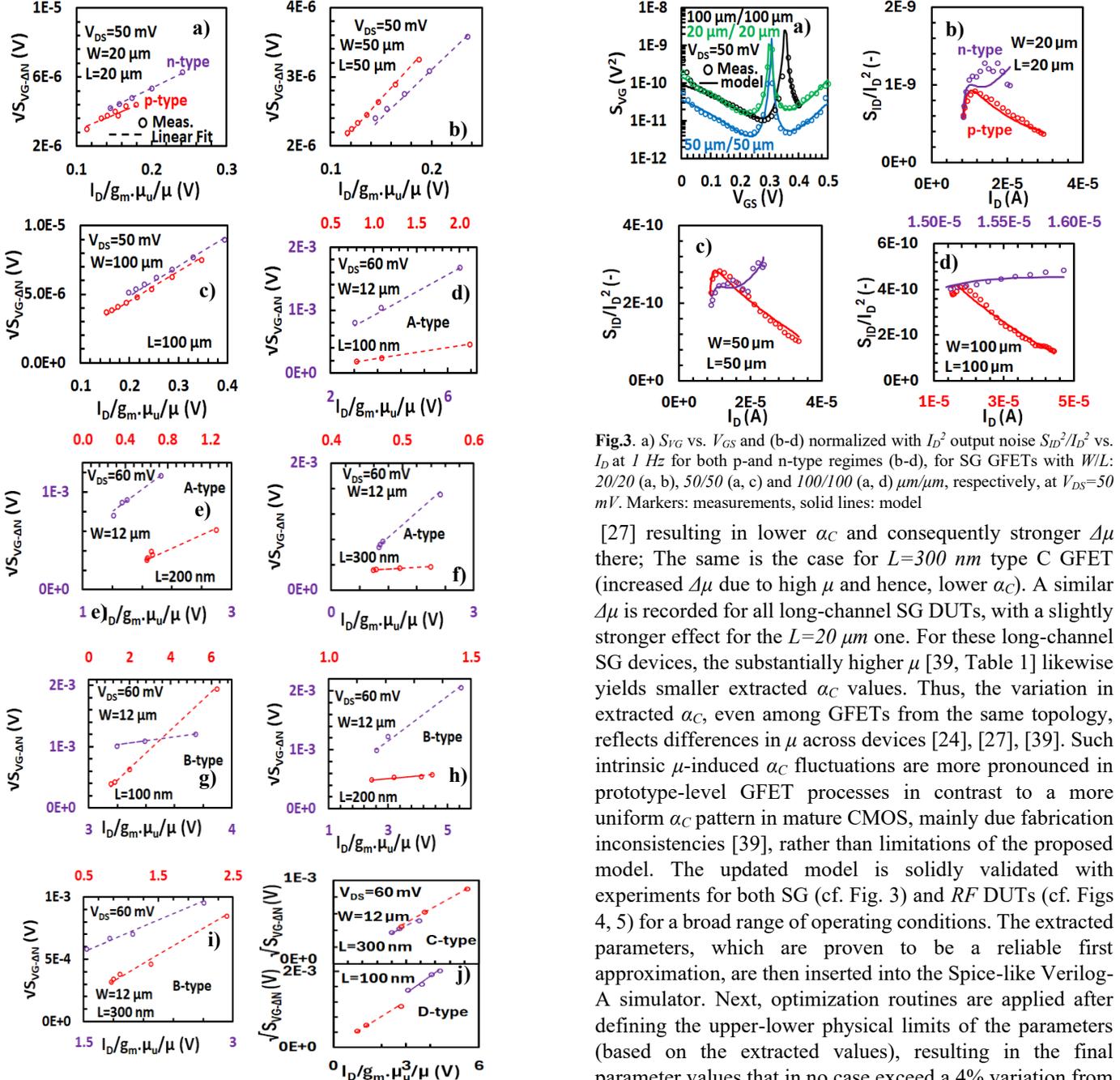

**Fig.2.** Square root of input noise $S_{VG}^{1/2}$ vs. $I_D/g_m.\mu_u/\mu$ at $(1\ Hz)^{1/2}$ for SG GFETs (a-c) with $W/L$=20/20 (a), 50/50 (b), 100/100 (c) $\mu m/\mu m$ and for type A (d-f), B (g-i), C (top panel of j) and D (bottom panel of j) *RF* GFETs with $W$=12 $\mu m$ and $L$=100 (d, g, j-bottom), 200 (e, h) and 300 (f, i, j- top) nm at $V_{DS}$=50 (a-c) and 60 (d-j) $mV$. Markers: measurements, dashed lines: linear fits. $I_D$, $\mu_u(\mu_u)$, are the drain current and the intrinsic low-field ($V_{GS}$-dependent degraded) mobility.

**Fig.3.** a) $S_{VG}$ vs. $V_{GS}$ and (b-d) normalized with $I_D^2$ output noise $S_{ID}^2/I_D^2$ vs. $I_D$ at *1 Hz* for both p-and n-type regimes (b-d), for SG GFETs with *W/L*: *20/20* (a, b), *50/50* (a, c) and *100/100* (a, d) $\mu m/\mu m$, respectively, at $V_{DS}$=50 $mV$. Markers: measurements, solid lines: model

[27] resulting in lower $\alpha_C$ and consequently stronger $\Delta\mu$ there; The same is the case for *L=300 nm* type C GFET (increased $\Delta\mu$ due to high $\mu$ and hence, lower $\alpha_C$). A similar $\Delta\mu$ is recorded for all long-channel SG DUTs, with a slightly stronger effect for the *L=20 μm* one. For these long-channel SG devices, the substantially higher $\mu$ [39, Table 1] likewise yields smaller extracted $\alpha_C$ values. Thus, the variation in extracted $\alpha_C$, even among GFETs from the same topology, reflects differences in $\mu$ across devices [24], [27], [39]. Such intrinsic $\mu$-induced $\alpha_C$ fluctuations are more pronounced in prototype-level GFET processes in contrast to a more uniform $\alpha_C$ pattern in mature CMOS, mainly due fabrication inconsistencies [39], rather than limitations of the proposed model. The updated model is solidly validated with experiments for both SG (cf. Fig. 3) and *RF* DUTs (cf. Figs 4, 5) for a broad range of operating conditions. The extracted parameters, which are proven to be a reliable first approximation, are then inserted into the Spice-like Verilog-A simulator. Next, optimization routines are applied after defining the upper-lower physical limits of the parameters (based on the extracted values), resulting in the final parameter values that in no case exceed a 4% variation from those obtained from the method. In this way, the physical validity of the parameters is ensured.

Both long and short channels from distinct GFET structures are included to verify that the model captures *LFN* across different device topologies. Notice that $S_{ID}/I_D^2$ in short-channel RF GFETs (cf. Fig. 5) is *1-2* orders of magnitude higher than in long-channel SG ones (cf. Fig. 3b-d), confirming its ~$1/WL$ dependence. Although separate parameter sets are used, analyzing these devices provides insight to develop more general and scalable models. Simulated lines solidly follow measured markers for both $S_{VG}$ vs. $V_{GS}$ (cf. Fig. 3a, Fig. 4) and $S_{ID}/I_D^2$ vs. $I_D$ (cf. Fig. 3b-d, Fig. 5). For $S_{ID}/I_D^2$, p(n)-type regions are presented separately with low to high $V_{DS}$ included in the analysis for the *RF* DUTs (cf. Fig. 5). The *VS*-induced reduction of *LFN* [19]-[20] is apparent at high $V_{DS}$ (right-panels). The analysis conducted here is extensive, including seven device structures, two distinct GFET architectures, and extended $V_{GS}$ and $V_{DS}$ sweeps; despite this complexity as well as the high sensitivity of *LFN* measurements especially for early-stage GFET technologies, the model consistently captures the main *LFN* trends for all cases, demonstrating generally good agreement across the full dataset. Overall, the fitting accuracy is consistent with prior GFET [16, Figs 2-3] and MOSFET [32, Fig. 2a], [35, Figs 4-6] studies, with average relative errors below ~10% (cf. Figs 3b-d, 5) and minor inconsistencies mainly arising from measurement fluctuations (cf. Fig. 4c-d). Fig. 5a additionally compares previous models: The previous version of the model [24] (dotted lines) reproduces the measurements decently but overvalues $N_t$ and $\alpha_H$ while it entails distinct $v_{sat}$ for *IV* and *LFN* fittings. In contrast, the proposed model achieves comparable accuracy but with parameters that remain physically consistent; notably, $\alpha_H$ aligns with CMOS case [30, Table 6.1], unlike prior GFET reports [19], [20], [24] where much higher values were required. Additionally, the simplified total *LFN* model (dashed lines), combining $\Delta N \sim g_m/I_D$ term [29, 31] with $\Delta\mu$, $\Delta R$ effects, undervalues *LFN* near the *CNP* while it fails to reproduce the VS-induced *LFN* reduction at high $V_{DS}$.

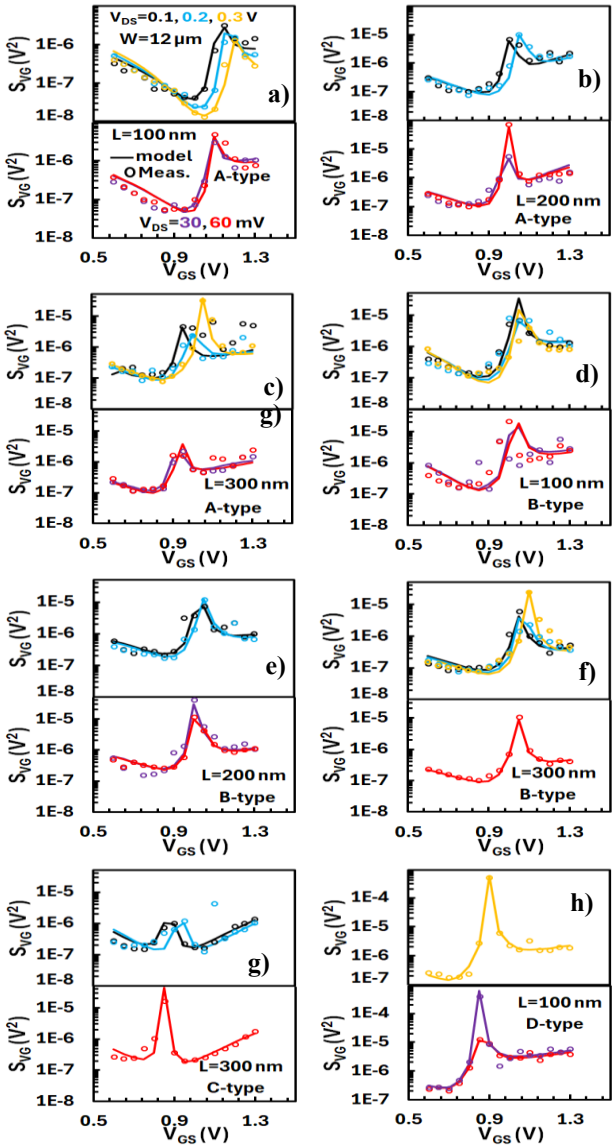

**Fig.4.** $S_{VG}$ vs. $V_{GS}$ at $V_{DS}=30, 60\ mV$ (bottom panels) and $V_{DS}=100, 200, 300\ mV$ (top panels), at $1\ Hz$ for type A (a-c), B (d-f), C (g) and D (h) *RF* GFETs with $W=12\ \mu m$ and $L=300$ (c, f, g), $200$ (b, e) and $100$ (a, d, h) *nm*. Markers: measurements, solid lines: model.

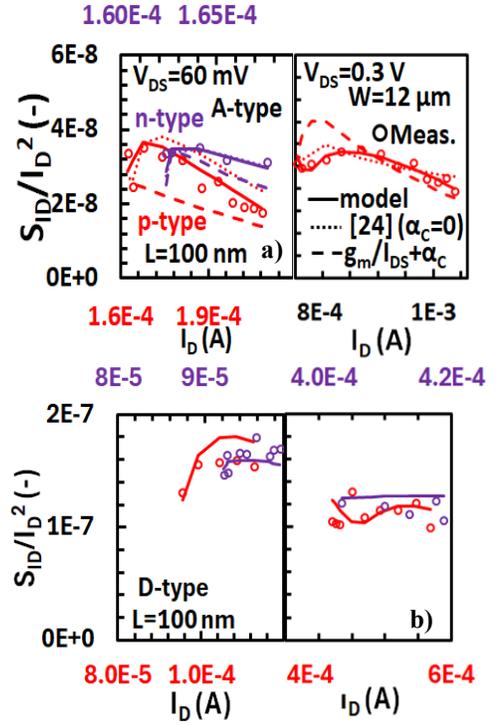

**Fig.5.** $S_{ID}^2/I_D^2$ vs. $I_D$ at $V_{DS}=60$ (left panels) and $300$ (right panels) *mV*, for both p(n)-type regimes, at $1\ Hz$ for type A (a) and D (b) *RF* GFETs with $W=12\ \mu m$ and $L=100\ nm$. Markers: measurements, solid lines: model, dotted lines: previous model [24] ($\alpha_C$ deactivated), dashed lines: simple model ($\Delta N \sim g_m/I_D + \alpha_C$).

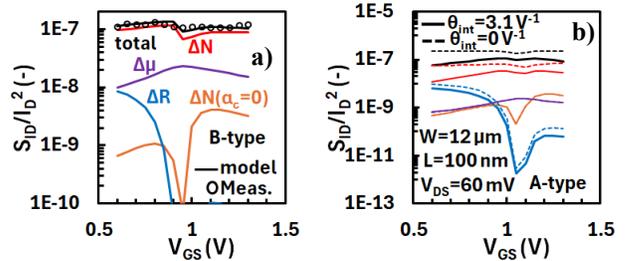

**Fig.6.** $S_{ID}^2/I_D^2$ vs. $V_{GS}$ at $1\ Hz$ for B (a) and A (b) -type *RF* GFETs with $W=12\ \mu m$ and $L=100\ nm$ at $V_{DS}=60\ mV$. Markers: measurements, solid lines: model, dashed lines in (b): $\theta_{int}=0\ V^{-1}$. Different colors represent different *LFN* contributions.

Fig. 6a depicts the different $S_{ID}/I_D^2$ contributions in addition to total model and measurements, vs. $V_{GS}$ at low $V_{DS}$

for the *L=100 nm* type B GFET; for low $V_{DS}$, global terms in (6a)-(9a) dominate over the *VS*-related ones in (6b)-(9b). *ΔN* prevails for every bias when the $α_C$ contribution is considered (*ΔNB+ΔNC*). *Δμ* presents a slight effect close to *CNP* while *ΔR* is strong solely at p-type region. With $α_C$=0, the physically consistent parameter extraction procedure proposed in this work results in a significantly lower *ΔN*, and therefore $N_t$ and $α_H$ must be overestimated to fit the experimental data, as done in previous works [24]. In Fig. 6b, $θ_{int}$ effect on *LFN* is examined for the *L=100 nm* type A DUT at the same low $V_{DS}$ as in Fig. 6a. A significant *ΔN* (and consequently total model) decrease is noticed when $θ_{int}$ is activated due to $μ_u$~$1/θ_{int}$ [39]. *ΔR* is slightly diminished with $θ_{int}$ especially at strong p(n)-type regions mainly through proportional $μ_u$ dependence of $g_m$, $g_{ds}$. Finally, *ΔNA=ΔN(α_C=0)*, and *Δμ* do not change with $θ_{int}$ as (6a) and (9a) do not contain $μ_u$.

## V. Conclusions

An improved physics-based GFET *LFN* model is formulated here to include for the first time correlated mobility fluctuations effect through Coulomb scattering, resulting in more intrinsically correct parameter values and ensuring the model's validity, accuracy, and scalability. A solid parameter extraction process adapted from CMOS technologies and updated to be compatible with GFETs features, is applied separately in both p(n)-type regions. In more detail, $N_t$, $α_c$ (*ΔN*) parameters are extracted via the linear fit of measured $\sqrt{S_{VG}}$ vs. $I_D/g_m·μ_u/μ$ at low $V_{DS}$ excluding high $I_D$ region where *ΔR* can be considerable. Next, $α_H$ (*Δμ*) can be extracted around *CNP* and then the method is repeated until optimum results are achieved. Finally, $S_{ΔR}$ (*ΔR*) is extracted at strong $I_D$. Precise modeling vs. measurements results for different types and footprints of DUTs covering all operating regions, are obtained. The physical essence of the final parameter values is ensured by extracting the four *LFN* parameters sequentially under physical constraints, which minimizes correlation and enhances the robustness of the methodology.